\documentclass[12pt]{iopart}
\makeatletter
\expandafter\let\csname equation*\endcsname\relax
\expandafter\let\csname endequation*\endcsname\relax
\makeatother
\usepackage{amsmath}
\usepackage{iopams}  
\usepackage{graphicx,subfigure}
\usepackage{ulem}
\usepackage{color}
\usepackage{times,txfonts}

\begin{document}

\title{Simulating cavity QED with spin-orbit coupled Bose-Einstein condensates revisited}

\author{Muhammad.~S.~Hasan$^1$ and Karol~Gietka$^2$}

\address{$^1$ Institute of Atomic and Molecular Sciences, Academia Sinica, Taipei, Taiwan}
\address{$^2$Institut f\"ur Theoretische Physik, Universit\"at Innsbruck, Technikerstra{\ss}e\,21a, A-6020 Innsbruck, Austria}

\ead{mshasan.official@gmail.com; karol.gietka@uibk.ac.at }

\begin{abstract}
Simulating cavity quantum electrodynamics in synthetic platforms offers a route to studying light--matter physics without real photons.  We reassess the frequently invoked analogy between harmonically trapped, spin--orbit-coupled atoms and the quantum Rabi and Dicke models, with particular emphasis on squeezing and entanglement.  A single non-interacting atom realizes the quantum Rabi Hamiltonian exactly, up to elementary spin and phase-space rotations.  For $N$ non-interacting atoms, however, the center-of-mass mode is only one member of a complete orthonormal set containing $N-1$ relative modes.  We prove that, after all modes are included, the spin--spin terms generated by the collective mode cancel exactly against those generated by the relative modes.  Equivalently, the original Hamiltonian is a sum of independent one-particle Hamiltonians and therefore cannot create interparticle entanglement from a separable state.  Contact interactions invalidate this cancellation but do not remove the relative modes or turn the exact many-body Hamiltonian into a Dicke model.  A one-mode Dicke description can arise only as a controlled low-energy approximation after all relevant relative excitations have been shown to be spectrally suppressed.  We discuss this distinction together with experimental scales, finite temperature, and possible mode-selection strategies.  We also distinguish the single-particle interference underlying density fringes from the interaction-stabilized stripe supersolid observed in many-body condensates.
\end{abstract}

\maketitle

\section{Introduction}
Cavity quantum electrodynamics (QED) investigates quantum light--matter interactions~\cite{Walther_2006_cavityQED,Ritsch2013cavityqed,Mivehvar02012021_QEDnew}.  Its characteristic phenomena include vacuum Rabi splitting~\cite{Eberly1983_VRS,Agarwal_1984_VRS,Kimble1992_VRS}, superradiance and subradiance~\cite{GROSS1982_superradiance,eberly1971superrad,kaiser2016_subradiance,hotter_2023_sub_sup}, and light-mediated entanglement~\cite{haroche_2001_entqed,thomposon_2011_vrs_sss,gietka_cond_enta_2025}.  Ultracold gases provide a complementary setting in which interactions, dimensionality, trapping, and synthetic gauge fields can be tuned~\cite{bloch2012ultracold}.  Raman-induced spin--orbit coupling changes the single-particle dispersion and, together with interactions, produces plane-wave, zero-momentum, and stripe phases, roton-like excitations, and unconventional superfluidity~\cite{Lin2009,busch2016reviewsocbec,Lingua:2018,Hasan:2022,Ghazaryan:2023,PhysRevLett.108.225301,PhysRevA.86.063621,Li2017,PhysRevLett.114.105301}.

The formal analogy with cavity QED originates at the one-particle level.  In a harmonic trap, the Raman spin--momentum coupling can be written exactly as a quantum Rabi interaction between one pseudospin and one quantized motional mode.  Hamner \emph{et al.}~\cite{socbecDicke2014} made the stronger many-body identification by using nonlinear interactions to motivate a single-condensate-orbital ansatz, replacing the $N$ motional coordinates by one harmonic amplitude, and interpreting the observed spin-polarization transition as the normal-to-superradiant transition of a Dicke model.  The subsequent review by Zhang \emph{et al.}~\cite{busch2016reviewsocbec} presents this construction in more detail.  After introducing collective spin operators, it substitutes one harmonic-oscillator operator for the atomic momenta and argues that interactions permit the restriction to this single mode because the atoms occupy the same condensate state.  This is a variational restriction of the many-body state space, not an operator identity for the full Hamiltonian.  Bose condensation into a common orbital does not eliminate the $N-1$ orthogonal relative coordinates.  Indeed, the same review notes that the single-mode step cannot be made in the absence of interactions.  The measured change of spin polarization and quasimomentum, and its Gross--Pitaevskii description, are not in dispute.  What is not exact is the identification of the underlying $N$-body Hamiltonian with a one-mode Dicke Hamiltonian.  Because the single-orbital projection removes the relative coordinates at the outset, agreement of mean-field order parameters with Dicke-model formulas does not by itself demonstrate a common quantum mediator, a Dicke superradiant transition, or cavity-like interparticle entanglement.

Lian \emph{et al.}~\cite{2013socsqueezing} and Huang and Hu~\cite{2015squeezingsocbec} subsequently used the same one-mode reduction to predict orbit-induced spin squeezing and spin--field squeezing.  After a unitary transformation, the retained oscillator generates an apparent collective $\hat S_z^2$ interaction, which was attributed to spin--orbit coupling.  However, both works also note that their single-spatial-mode approximation is invalid in the absence of atom--atom interactions.  This exposes the central inconsistency.  Interactions are invoked to lock the atoms into one orbital, but the collective squeezing term is then interpreted as being generated by spin--orbit coupling alone.  For genuinely non-interacting atoms the microscopic Hamiltonian is a sum of independent one-particle Hamiltonians, and local spin--motion coupling cannot create interparticle entanglement.

Chen, Zhang, and Pu~\cite{PhysRevA.102.023317} identified this conceptual problem, stating that the earlier Dicke-like mapping with the interaction term neglected relied on unjustified approximations.  They instead projected an interacting gas onto two dressed momentum modes and obtained a one-axis-twisting coefficient directly from the contact interaction.  In their treatment the interaction is essential for squeezing, while spin--orbit coupling tunes its strength.  Their analysis therefore supplies the correct interacting mechanism, but it does not identify the omitted harmonic coordinates or show explicitly how the spurious collective term disappears when the full non-interacting motional Hilbert space is restored.

That missing mode accounting is the issue resolved here.  Keeping only the center-of-mass mode produces a Dicke Hamiltonian, but a complete orthonormal transformation also produces $N-1$ relative modes coupled to non-collective spin combinations.  We show analytically that, in the non-interacting harmonic model, the effective spin--spin terms from the complete set of modes cancel exactly.  This is also immediate in particle coordinates, where the Hamiltonian factorizes into independent one-particle quantum Rabi Hamiltonians.  The earlier single-mode phase-transition and squeezing predictions must therefore be read as projected mean-field or interacting models, not as consequences of the unprojected non-interacting Hamiltonian.  Contact interactions remove the factorization and can permit genuine correlations, but they do not eliminate the relative modes or establish an exact Dicke mapping.  At most, a Dicke Hamiltonian may emerge as an effective low-energy model when a quantitative spectral hierarchy suppresses every relevant relative excitation.  Without that additional demonstration the interacting SOC condensate remains a multimode system.

This distinction is also essential for the stripe phase.  A coherent superposition of the two minima of the single-particle dispersion produces density fringes already at the mean-field one-body level.  The selection, stability, collective excitations, and supersolid properties of the equilibrium stripe phase nevertheless depend on interatomic interactions~\cite{Li2017,PhysRevLett.110.235302,Ji2014FiniteTemperature}.  We therefore distinguish throughout between single-particle interference and an interaction-stabilized many-body stripe phase.

The manuscript is organized as follows.  Section~\ref{Sec:SOCBEC_QRM} establishes the one-particle quantum Rabi mapping and discusses its domain of validity.  Section~\ref{Sec:Dicke_model} summarizes the collective Dicke benchmark.  Sections~\ref{Sec:two_atoms}--\ref{Sec:N_atoms} expose the collective and relative couplings and prove their exact cancellation in the non-interacting model.  Section~\ref{Sec:interactions} discusses contact interactions, experimental scales, temperature, anisotropy, and mitigation strategies.  We conclude in Sec.~\ref{Sec:Conclusion}.

\section{Spin--orbit coupled atom in a harmonic trap as a quantum Rabi model} \label{Sec:SOCBEC_QRM}
We begin by revisiting the quantum Rabi model~\cite{Xie_2017}, which describes a single two-level atom coupled to a single mode of the quantized electromagnetic field (throughout the paper, we set $\hbar = 1$ for convenience)
\begin{align}\label{eq:rabi}
    \hat H_{QR} = \omega \hat a^\dagger \hat a + \frac{\Omega}{2} \hat \sigma_z + \frac{g}{2} \left(\hat a^\dagger + \hat a\right) \hat \sigma_x,
\end{align}
where $\hat a$ ($\hat a^\dagger$) annihilates (creates) an excitation (photon) of energy $\omega$, $\hat \sigma_\alpha$ are the Pauli matrices describing the atomic internal states with transition frequency $\Omega$, and $g$ is the light--matter coupling strength.

By contrast, the Hamiltonian for a spin-orbit coupled two-level atom in a harmonic trap takes the form~\cite{socbec2011,socbecDicke2014}
\begin{align}\label{eq:H1}
    \hat H_{SOC} = \frac{(\hat p + k \hat \sigma_z)^2}{2m} + \frac{1}{2} m \omega^2 \hat x^2 + \frac{\Omega}{2} \hat \sigma_x,
\end{align}
where $\hat x$ and $\hat p = - i \partial_x$ are the position and momentum operators, $m$ is the atomic mass, $\omega$ is the trap frequency, $\Omega$ is the two-photon Raman coupling strength, and $k$ characterizes the spin-orbit coupling strength (typically associated with recoil momentum).  To draw an explicit analogy with cavity QED systems, we perform a spin rotation around the $y$-axis and express Eq.~\eqref{eq:H1} in terms of the harmonic oscillator annihilation and creation operators with the help of an additional phase-space rotation. The resulting spin-orbit coupling Hamiltonian becomes
\begin{align}\label{eq:rabi_mapping}
    \hat H = \omega \hat a^\dagger \hat a + \frac{\Omega}{2} \hat \sigma_z + k \sqrt{\frac{\omega}{2m}} \left(\hat a + \hat a^\dagger\right) \hat \sigma_x.
\end{align}
This is formally equivalent to the quantum Rabi model Hamiltonian $H_{QR}$ with the identification
\begin{align}
\frac{g}{2} = k \sqrt{\frac{\omega}{2m}}.
\end{align}
This mapping enables the exploration of light--matter interaction physics in a photon-free setting, where quantized motional excitations take the role of photons.  A key advantage is access to the ultrastrong-coupling regime~\cite{RevModPhys.91.025005,usc2019miran,QIN20241}, in which $g$ is comparable to the bare frequencies $\omega$ and $\Omega$.  Vacuum squeezing~\cite{PhysRevA.110.063703}, virtual excitations~\cite{liberato2017GSvirt}, and criticality-induced entanglement~\cite{usc2019miran} then become prominent.  The mapping in Eqs.~\eqref{eq:H1}--\eqref{eq:rabi_mapping} is exact for one atom.  Its extension to $N$ atoms is the question examined below.

\begin{figure}[htb!]
    \centering
    \includegraphics[width=0.45\textwidth]{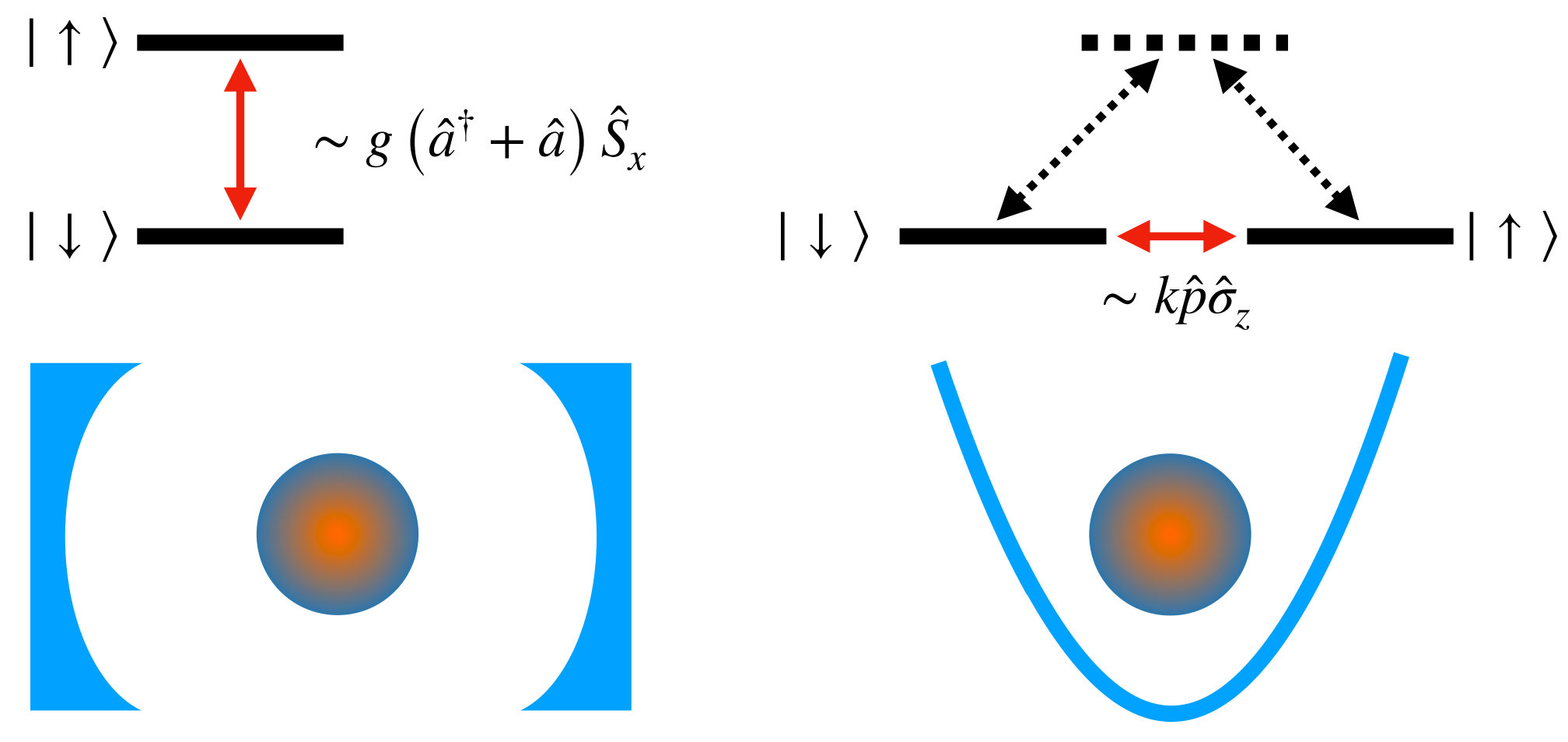}
    \caption{(a) The Dicke model couples an ensemble of two-level atoms to one cavity mode.  (b) Raman beams couple two internal states of harmonically trapped atoms.  For one atom, the two couplings are unitarily equivalent.  For $N$ atoms, the cavity interaction remains collective whereas the spin--orbit Hamiltonian contains both a collective center-of-mass coupling and $N-1$ relative-mode couplings.
}
    \label{fig:scheme}
\end{figure}

\subsection{Squeezed virtual excitations}\label{Sec: Virtual_excitations}
We now consider the strongly off-resonant regime, $\omega \ll \Omega$, where the atomic dynamics can be adiabatically eliminated~\cite{PhysRevLett.115.180404}. Expanding the Hamiltonian \eqref{eq:rabi} up to quadratic order yields an effective model for the bosonic mode
\begin{align}\label{eq:invertedH}
\hat{H} \approx \omega \hat{a}^\dagger \hat{a} - \frac{g^2}{4\Omega} \left(\hat{a} + \hat{a}^{\dagger} \right)^2.
\end{align}
This describes a squeezed harmonic oscillator whose eigenstates are squeezed Fock states
\begin{align} \label{eq:squeezeoperator}
\hat{S}(\xi) |n\rangle = \exp\left[\frac{\xi}{2} \left(\hat{a}^2 - \hat{a}^{\dagger 2}\right)\right] |n\rangle,
\end{align}
where $\hat{S}(\xi)$ is the squeezing operator and
\begin{align}
\xi = -\frac{1}{4} \log\left(1 - \frac{g^2}{\omega \Omega}\right)
\end{align}
is the squeezing parameter. Squeezing redistributes quantum fluctuations between conjugate quadratures~\cite{Loudon01061987}, modifying the uncertainty relations of the field. In the vacuum (or any coherent) state, the dimensionless quadrature operators
\begin{align}
\hat{x} = \frac{1}{\sqrt{2}}(\hat{a} + \hat{a}^\dagger), \qquad
\hat{p} = \frac{1}{\sqrt{2}i}(\hat{a} - \hat{a}^\dagger)
\end{align}
have equal variances, $\Delta^2 \hat{x} = \Delta^2 \hat{p} = 1/2$, and the corresponding Wigner function is a circular Gaussian. In contrast, the squeezed vacuum state $\hat{S}(\xi) |0\rangle$ exhibits anisotropic fluctuations
\begin{align}
\Delta^2 \hat{x} = \frac{1}{2} e^{2\xi}, \qquad
\Delta^2 \hat{p} = \frac{1}{2} e^{-2\xi}.
\end{align}
The uncertainty in one quadrature is suppressed while the other is amplified, preserving the Heisenberg relation
\begin{align}
\Delta^2 \hat{x} \, \Delta^2 \hat{p} = \frac{1}{4}.
\end{align}
In phase space, this corresponds to an elliptical Wigner distribution---the geometric signature of squeezing.

Importantly, the squeezed ground state contains a non-zero number of excitations
\begin{align}
\langle \hat{a}^\dagger \hat{a} \rangle = \sinh^2 \xi \geq 0,
\end{align}
suggesting the presence of photons even in the ground state. However, this does not imply real radiation. The system remains in its lowest energy eigenstate and cannot emit energy. The resolution lies in recognizing that transition operators in a consistent open-system treatment connect dressed eigenstates rather than creating bare excitations~\cite{PhysRevA.84.043832,stefanini2026lindbladme,10.21468/SciPostPhysLectNotes.68}.  Defining the normal mode
\begin{align}
\hat{c} = \hat{S}^\dagger(\xi) \hat{a} \hat{S}(\xi), \qquad
\hat{c}^\dagger = \hat{S}^\dagger(\xi) \hat{a}^\dagger \hat{S}(\xi),
\end{align}
in terms of which the Hamiltonian takes the diagonal form
\begin{align}
\hat{H} = \omega e^{-2\xi} \hat{c}^\dagger \hat{c} = \omega \sqrt{1 - \frac{g^2}{\omega \Omega}} \, \hat{c}^\dagger \hat{c}.
\end{align}
Thus, the dressed ground state is the vacuum of $\hat c$ but contains a non-zero number of excitations with respect to the uncoupled operator $\hat a$.  The latter are conventionally called \emph{virtual photons}.  Their number is representation dependent, while observable consequences such as the normal-mode frequency shift and quench-induced excitations are not.

Observing virtual photons directly in cavity QED remains challenging~\cite{RevModPhys.91.025005,PhysRevResearch.6.013008}.  In the one-atom simulator, one could instead prepare the dressed ground state and quench the spin--orbit coupling.  The resulting real motional excitations and quadrature anisotropy could be measured by time-of-flight imaging at different trap phases.

\subsection{Critical Rabi-model limit and stripe interference} \label{Sec: Superradiant_PT}
The quadratic Hamiltonian~\eqref{eq:invertedH} softens at
\begin{align}
g_c \equiv \sqrt{\omega \Omega}
\end{align}
and becomes unstable above this value~\cite{PhysRevE.104.034132}.  The full quantum Rabi model remains bounded.  A genuine quantum phase transition arises only in the scaling limit $\Omega/\omega\rightarrow\infty$.  At finite frequency ratio there is a sharp crossover rather than a thermodynamic phase transition~\cite{PhysRevLett.115.180404}.  Deep in the strong-coupling regime the lowest parity eigenstate approaches the light--matter cat state
\begin{align}
    |\Psi_0\rangle = \frac{|-\alpha\rangle \otimes |+\rangle - |\alpha\rangle \otimes |-\rangle}{\sqrt{2}},
\end{align}
where $|\pm\alpha\rangle$ are oppositely displaced oscillator states and $|\pm\rangle$ are spin states~\cite{stassi2021unveilingveilingschrodingercat,usc2019miran}.  In electric-dipole cavity QED, the equilibrium Dicke transition predicted by a truncated model may be forbidden by the diamagnetic $A^2$ term and the Thomas--Reiche--Kuhn sum rule~\cite{Rzazewski1975A2,Bialynicki1979NoGo,Nataf2010NoGo,Viehmann2011NoGo}.  Raman-engineered spin--orbit coupling is an effective Hamiltonian and is not constrained by the same microscopic electromagnetic sum rule.  The distinction must nevertheless be made when the simulator is compared with a physical cavity.

For a non-interacting Raman-coupled atom, the two minima of the lower dispersion merge at
\begin{align}
\Omega_c \equiv \frac{2k^2}{m}.
\end{align}
Below this value, a coherent superposition of opposite-momentum minima produces one-body density fringes.  This interference mechanism does not by itself establish a thermodynamic stripe phase or many-body entanglement.  Without interactions, plane-wave and superposition states may be degenerate.  In a condensate, interactions select among the stripe, plane-wave, and zero-momentum phases and determine the stripe contrast, stability, collective modes, and supersolid response~\cite{Li2017,PhysRevLett.110.235302,Ji2014FiniteTemperature}.  We therefore use ``stripe interference'' for the one-body effect and reserve ``stripe phase'' for the interaction-stabilized many-body state.

\section{Many-body case: Dicke Model}\label{Sec:Dicke_model}
The many-body counterpart of the quantum Rabi model is the Dicke model~\cite{10.1098/rsta.2010.0333}, in which one resonator mode couples to an entire atomic ensemble,
\begin{align}
    \hat H = \omega \hat a^\dagger \hat a + {\Omega}\hat S_z +\frac{g}{\sqrt{N}}\left(\hat a^\dagger + \hat a\right) \hat S_x,
\end{align}
where $\hat S_\alpha = \frac{1}{2}\sum_{i=1}^N \hat \sigma_\alpha^{(i)}$ are collective spin operators.  A common bosonic mode can mediate effective all-to-all interactions~\cite{Mivehvar02012021_QEDnew}.  We briefly recall three useful frequency regimes.

\subsection{$\Omega\ll\omega$ regime}\label{Sec:LMG_model}
For clarity, we now consider the regime where the cavity dynamics is much faster than the atom dynamics, allowing the cavity mode to be adiabatically eliminated from the description. The effective Hamiltonian then becomes~\cite{PhysRevA.85.043821}
\begin{align}
\hat{H} = \Omega \hat{S}_z - \frac{g^2}{N \omega} \hat{S}_x^2,
\end{align}
where the all-to-all interaction term \(\hat{S}_x^2\) reflects the induced dipole-dipole interactions arising from the elimination of the cavity mode. The above Hamiltonian is known as the Lipkin-Meshkov-Glick model~\cite{PhysRevA.85.043821}, a many-body spin Hamiltonian known to exhibit spin-squeezing and phase transition where the ground state is a maximally entangled atomic cat state~\cite{PhysRevLett.99.050402}. Spin squeezing is closely analogous to the bosonic squeezing discussed earlier~\cite{MA201189}. In the same way that the squeezing operator \(\hat{S}(\xi)\) deforms the uncertainty ellipse in phase space, the nonlinear interaction term \(\hat{S}_x^2\)---also known as one-axis twisting~\cite{PhysRevA.47.5138}---deforms the uncertainty region of the collective spin state on the collective Bloch sphere. For an uncorrelated spin-coherent state, quantum fluctuations are evenly distributed in the transverse plane perpendicular to the mean spin direction. The \(\hat{S}_x^2\) interaction shears (twists) this distribution, leading to reduced fluctuations in one transverse direction (spin squeezing), and increased fluctuations in the conjugate direction, analogous to the squeezing of \(\hat{x}\) and \(\hat{p}\) quadratures of a harmonic oscillator. 

Metrologically useful spin squeezing is quantified by the Wineland parameter~\cite{PhysRevA.46.R6797}
\begin{align}
    \xi_R^2 = \frac{N\min_{\boldsymbol n_\perp}(\Delta \hat S_{\boldsymbol n_\perp})^2}{|\langle\hat{\boldsymbol S}\rangle|^2},
\end{align}
where the minimization is over directions orthogonal to the mean spin.  The condition $\xi_R^2<1$ witnesses metrologically useful entanglement.  If the mean spin remains close to $-N\boldsymbol e_z/2$, this reduces to
\begin{align}
    \xi_R^2 \simeq \frac{4\min_{\perp}(\Delta \hat S_{\perp})^2}{N},
\end{align}
but below we evaluate the full expression, including the mean-spin denominator.

\subsection{$\Omega \approx \omega$ regime}
In the resonant case, where the atomic transition frequency matches the resonator frequency, it is convenient to apply the Holstein-Primakoff transformation along with the large atom number approximation~\cite{PhysRevE.67.066203}. Under these conditions, the Dicke Hamiltonian reduces to
\[
\hat H = \omega \hat a^\dagger \hat a + \Omega \hat b^\dagger \hat b + \frac{g}{2}(\hat a + \hat a^\dagger)(\hat b + \hat b^\dagger),
\]
describing two coupled harmonic oscillators. The ground state of this system is an entangled two-mode squeezed vacuum~\cite{mirkhalaf2025frequencyshiftsheraldingground}. However, observing this entanglement directly in cavity QED experiments is challenging due to the way these systems couple to their environment. Specifically, cavity QED systems typically couple to the environment through their eigenmodes, making the system effectively appear as two independent harmonic oscillators from the outside.
\[
\hat H = \omega \exp(-2\xi_-)\,\hat c^\dagger \hat c + \omega \exp(-2\xi_+)\,\hat d^\dagger \hat d,
\]
where the two-mode squeezing present in the ground state, characterized by the squeezing parameters
\[
\xi_\pm = -\frac{1}{4}\log(1 \pm g/g_c),
\]
manifests as modified frequencies of the new normal modes represented by \(\hat c\) and \(\hat d\)~\cite{PhysRevA.110.063703}.
\begin{align}
     \omega \exp(-2\xi_\pm)
\end{align}

If it were possible to simulate the Dicke model using spin-orbit coupled Bose-Einstein condensates, it would be much easier to observe these quantum phenomena. This is because the center-of-mass mode in ultracold atomic systems is far better isolated from environmental decoherence than photons in a cavity, and there is significantly more control over the trapping potential, which plays the role of the harmonic oscillator in this context. 

\subsection{$\Omega \gg \omega$ regime}
Finally, in the regime which we already looked at in the section about the quantum Rabi model, the spin degree of freedom can be eliminated due to the large energy separation \(\Omega \gg \omega\), allowing for an adiabatic elimination of the atomic (spin) dynamics. This leads to the effective bosonic Hamiltonian
\begin{align}
\hat{H} \approx \omega \hat{a}^\dagger \hat{a} - \frac{g^2}{4\Omega}\left(\hat{a} + \hat{a}^{\dagger}\right)^2,
\end{align}
which is identical to the one obtained in the strongly off-resonant quantum Rabi model. This result underscores the universality of the underlying physics. Both systems exhibit the same Gaussian squeezing despite originating from different microscopic interactions, single atom light-matter coupling in one case and collective spin-cavity interactions in the other.

In the next sections, we revisit the mapping between the Dicke model and spin-orbit coupled Bose-Einstein condensates, where this effective description plays a key role in understanding emergent analogies.

\section{Two atoms in a harmonic trap}\label{Sec:two_atoms}
We begin with two non-interacting atoms,
\begin{align}\label{eq:socbecH}
    \hat H = \sum_{i=1}^2 \left[\omega \hat a_i^\dagger \hat a_i + \frac{\Omega}{2} \hat \sigma_z^{(i)} + \frac{g}{2} \hat \sigma_x^{(i)} \left( \hat a_i + \hat a_i^\dagger \right)\right].
\end{align}
Unlike the Dicke Hamiltonian, Eq.~\eqref{eq:socbecH} contains one motional mode per atom.  Defining the center-of-mass and relative modes
\begin{align}
    \hat c = \frac{\hat a_1 + \hat a_2}{\sqrt{2}}, 
    \quad
    \hat d = \frac{\hat a_1 - \hat a_2}{\sqrt{2}},
    \label{com_rel_ops}
\end{align}
and substituting their inverse relations gives
\begin{align}
    \hat a_1 = \frac{\hat c + \hat d}{\sqrt{2}}, 
    \quad
    \hat a_2 = \frac{\hat c - \hat d}{\sqrt{2}}.
\end{align}
\begin{align}
    \hat H =& \, \omega \hat c^\dagger \hat c 
    + \frac{\Omega}{2} \left( \hat \sigma_z^{(1)} + \hat \sigma_z^{(2)} \right)
    + \frac{g}{2\sqrt{2}} \left( \hat \sigma_x^{(1)} + \hat \sigma_x^{(2)} \right) \left( \hat c + \hat c^\dagger \right) \nonumber \\
    &+ \omega \hat d^\dagger \hat d 
    + \frac{g}{2\sqrt{2}} \left( \hat \sigma_x^{(1)} - \hat \sigma_x^{(2)} \right) \left( \hat d + \hat d^\dagger \right),
    \label{com_rel_H}
\end{align}
The first line contains the desired collective Dicke coupling, but the second line is not optional.  In the dispersive regime $\Omega\ll\omega$, eliminating both oscillators gives
\begin{align}
\hat H_{\rm eff}^{(2)}={}&\frac{\Omega}{2}(\hat\sigma_z^{(1)}+\hat\sigma_z^{(2)})
-\frac{g^2}{8\omega}(\hat\sigma_x^{(1)}+\hat\sigma_x^{(2)})^2\nonumber\\
&-\frac{g^2}{8\omega}(\hat\sigma_x^{(1)}-\hat\sigma_x^{(2)})^2\nonumber\\
={}&\frac{\Omega}{2}(\hat\sigma_z^{(1)}+\hat\sigma_z^{(2)})-\frac{g^2}{2\omega}\mathbb I.
\end{align}
The cross term $\hat\sigma_x^{(1)}\hat\sigma_x^{(2)}$ generated by the collective mode is cancelled exactly by the relative mode.  The cancellation is algebraic rather than a dephasing approximation.

\section{Three atoms in a harmonic trap}\label{Sec:three_atoms}
For three atoms the correctly summed Hamiltonian is
\begin{align}\label{eq:H_three_atoms}
    \hat H = \sum_{i=1}^3\left[\omega \hat a_i^\dagger \hat a_i + \frac{\Omega}{2} \hat \sigma_z^{(i)} + \frac{g}{2}\hat \sigma_x^{(i)}\left(\hat a_i + \hat a_i^\dagger\right)\right].
\end{align}
The Jacobi transformation~\cite{Amico:2014}
\begin{align}
\begin{split}
    \hat c &= \frac{1}{\sqrt{3}} (\hat a_1+\hat a_2+\hat a_3), \\
    \hat d &= \frac{1}{\sqrt{2}}(\hat a_1 - \hat a_2), \\
    \hat f &= \frac{1}{\sqrt{6}}(-\hat a_1-\hat a_2+2 \hat a_3),
\end{split}
\end{align}
contains the center-of-mass mode $\hat c$ and two relative modes $\hat d,\hat f$.  It is orthonormal and preserves the canonical commutators.  Its inverse is
\begin{align}
\begin{split}\label{jacobi_ops}
   \hat a_1 &= \frac{1}{\sqrt{3}} \hat c + \frac{1}{\sqrt{2}} \hat d -\frac{1}{\sqrt{6}} \hat f, \\
   \hat a_2 &= \frac{1}{\sqrt{3}} \hat c - \frac{1}{\sqrt{2}} \hat d -\frac{1}{\sqrt{6}} \hat f, \\
   \hat a_3 &= \frac{1}{\sqrt{3}} \hat c + \frac{\sqrt{2}}{\sqrt{3}} \hat f.
\end{split}
\end{align}
Substituting these into Eq.~\eqref{eq:H_three_atoms}, the Hamiltonian becomes
\begin{align}   
\begin{split}
    \hat H = &~\omega \left(\hat{c}^\dagger \hat c + \hat{d}^\dagger \hat d + \hat{f}^\dagger \hat f\right) + \frac{\Omega}{2}\sum_{i=1}^3 \hat \sigma_z^{(i)}  \\
    &+ \frac{g}{2\sqrt{3}}\left( \hat{c}^\dagger + \hat c\right) \left( \hat\sigma_x^{(1)} + \hat \sigma_x^{(2)} + \hat \sigma_x^{(3)} \right) \\
    &+ \frac{g}{2\sqrt{2}}\left( \hat{d}^\dagger + \hat d\right) \left( \hat \sigma_x^{(1)} - \hat \sigma_x^{(2)}\right)  \\ 
    &+ \frac{g}{2\sqrt{6}}\left( \hat{f}^\dagger + \hat f\right) \left( -\hat\sigma_x^{(1)} - \hat \sigma_x^{(2)} + 2 \hat \sigma_x^{(3)} \right).  \label{three_atoms_collective_relative}
\end{split}
\end{align}
The second line of Eq.~\eqref{three_atoms_collective_relative} is collective, whereas the final two lines couple relative motion to orthogonal spin combinations.  Eliminating only the collective mode produces squeezing.  Eliminating the complete set again gives only a constant, as follows from the general proof below.

\section{Exact result for $N$ non-interacting atoms}\label{Sec:N_atoms}
Let $O_{\mu i}$ be any real orthogonal matrix whose first row is $O_{0i}=1/\sqrt N$, and define $\hat b_\mu=\sum_iO_{\mu i}\hat a_i$.  The first mode $\hat b_0\equiv\hat P$ is the center-of-mass mode and the remaining modes are relative modes.  The $N$-atom Hamiltonian becomes
\begin{align}\label{eq:general_jacobi_H}
\hat H={}&\omega\sum_{\mu=0}^{N-1}\hat b_\mu^\dagger\hat b_\mu+\frac{\Omega}{2}\sum_{i=1}^N\hat\sigma_z^{(i)}\nonumber\\
&+\frac g2\sum_{\mu=0}^{N-1}(\hat b_\mu+\hat b_\mu^\dagger)
\left(\sum_{i=1}^NO_{\mu i}\hat\sigma_x^{(i)}\right).
\end{align}
The $\mu=0$ term is precisely the collective Dicke coupling.  It is accompanied, with fixed strengths, by all $N-1$ orthogonal relative-mode couplings.

For $\Omega\ll\omega$, eliminating the complete set of modes yields
\begin{align}\label{eq:all_modes_effective}
\hat H_{\rm eff}={}&\frac{\Omega}{2}\sum_i\hat\sigma_z^{(i)}
-\frac{g^2}{4\omega}\sum_{\mu=0}^{N-1}
\left(\sum_iO_{\mu i}\hat\sigma_x^{(i)}\right)^2.
\end{align}
Orthogonal completeness, $\sum_\mu O_{\mu i}O_{\mu j}=\delta_{ij}$, gives
\begin{align}\label{eq:cancellation_identity}
\sum_\mu\left(\sum_iO_{\mu i}\hat\sigma_x^{(i)}\right)^2
=\sum_i(\hat\sigma_x^{(i)})^2=N\mathbb I.
\end{align}
All induced spin--spin terms therefore cancel exactly.  This proof also shows why discarding the relative modes produces a spurious collective interaction.

The absence of interparticle entanglement is not restricted to the dispersive approximation.  In particle coordinates the Hamiltonian is exactly $\hat H=\sum_i\hat h_i$, with $[\hat h_i,\hat h_j]=0$ for $i\ne j$, and hence $\hat U(t)=\bigotimes_i\exp(-i\hat h_it)$.  A separable internal--motional state remains separable between particles.  For identical non-interacting bosons initially occupying one spinor orbital, the symmetrized Hartree product likewise remains a Hartree product.  Local spin--motion entanglement may be generated, but genuine interparticle entanglement and collective spin squeezing cannot.

\begin{figure*}[t]
\centering
\includegraphics[width=\textwidth]{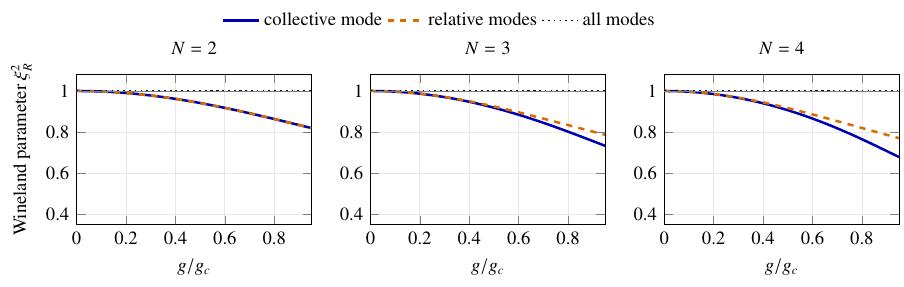}
 \caption{Ground-state Wineland parameter in the dispersive spin model for $N=2,3,4$.  The collective mode alone and the complete relative-mode subset separately produce squeezing, but the full orthonormal mode set gives $\xi_R^2=1$ because of Eq.~\eqref{eq:cancellation_identity}.}
\label{fig:wineland}
\end{figure*}

Figure~\ref{fig:wineland} evaluates the full Wineland parameter for the ground state of Eq.~\eqref{eq:all_modes_effective}.  Retaining only the collective mode or only the relative subset produces squeezing along different transverse axes.  Including the complete orthonormal set restores $\xi_R^2=1$ exactly.  The calculation uses the collective-spin convention $\hat S_\alpha=\sum_i\hat\sigma_\alpha^{(i)}/2$ and does not invoke a large-$N$ approximation.

In the complementary regime $\Omega\gg\omega$, eliminating each spin gives
\begin{align}
\hat{H} \approx \sum_{i=1}^N \left[ \omega \hat{a}^\dagger_i \hat{a}_i - \frac{g^2}{4\Omega} \left( \hat{a}_i + \hat{a}^\dagger_i \right)^2 \right].
\end{align}
This is a sum of independently squeezed one-particle oscillators.  Motional squeezing is present, but it is not a collective enhancement and does not create interparticle correlations.

\section{Interactions and experimental considerations}\label{Sec:interactions}
\subsection{Contact interactions}
For a two-component gas, the one-dimensional contact interaction can be written
\begin{align}\label{eq:contact}
\hat H_{\rm int}=\sum_{i<j}\delta(\hat x_i-\hat x_j)\big[&g_{\uparrow\uparrow}\hat P_\uparrow^{(i)}\hat P_\uparrow^{(j)}
+g_{\downarrow\downarrow}\hat P_\downarrow^{(i)}\hat P_\downarrow^{(j)}\nonumber\\
&+g_{\uparrow\downarrow}(\hat P_\uparrow^{(i)}\hat P_\downarrow^{(j)}+\hat P_\downarrow^{(i)}\hat P_\uparrow^{(j)})\big],
\end{align}
where $\hat P_s$ projects onto component $s$.  Since Eq.~\eqref{eq:contact} depends only on coordinate differences, a spin-independent interaction leaves the bare center-of-mass oscillator unchanged in a harmonic trap but changes the energies and wavefunctions of the relative modes.  The equal-frequency completeness step leading to Eq.~\ref{eq:cancellation_identity} then no longer applies to the energy denominators and matrix elements.  Importantly, this breakdown of the cancellation does not eliminate any motional degrees of freedom.  The spin--orbit term still couples the spins to a tower of interaction-dressed relative excitations.  The exact interacting Hamiltonian is therefore generically multimode, not a Dicke Hamiltonian.

The effect has no universal sign.  If an interaction shift $\Delta_{\rm rel}$ makes all relevant relative transitions far off resonance, $|\Delta_{\rm rel}|$ much larger than their spin--orbit matrix elements, virtual population of the relative modes is suppressed and the collective contribution can dominate.  This yields only a controlled approximate one-mode description, and requires the separation to hold for the complete set of relative transitions retained at the relevant energy and temperature.  Near an interaction-shifted relative resonance, the same modes are enhanced and produce faster loss of collective spin contrast.  Differences among $g_{\uparrow\uparrow}$, $g_{\downarrow\downarrow}$, and $g_{\uparrow\downarrow}$ additionally create spin-dependent nonlinearities that can themselves generate squeezing, as analyzed in Ref.~\cite{PhysRevA.102.023317}.  Interactions may therefore improve or degrade collective behavior depending on detuning, density, and scattering lengths.  Equation~\eqref{eq:cancellation_identity} is a theorem for the non-interacting harmonic model, while its failure in an interacting condensate is not by itself evidence for a Dicke model.

\subsection{Representative scales}
As a concrete benchmark, the $^{87}$Rb experiment of Lin \emph{et al.}~\cite{socbec2011} used trap frequencies $(f_x,f_y,f_z)\simeq(50,50,140)$~Hz and Raman recoil near $E_R/h\simeq3.55$~kHz.  The associated scales are summarized in Table~\ref{tab:scales}.
\begin{table}[b]
\caption{Representative $^{87}$Rb scales for Raman spin--orbit coupling.  Frequencies in the Hamiltonian are angular frequencies.}
\label{tab:scales}
\begin{indented}
\item[]\begin{tabular}{@{}lc}
\br
Quantity & Representative value\\
\mr
Axial trap frequency $\omega_x/2\pi$ & $50$ Hz\\
Trap response time $1/\omega_x$ & $3.2$ ms\\
Trap period $2\pi/\omega_x$ & $20$ ms\\
Recoil frequency $E_R/h$ & $3.55$ kHz\\
Recoil time $h/E_R$ & $0.28$ ms\\
Raman scale $\hbar\Omega=0$--$4E_R$ & $\Omega/2\pi=0$--$14.2$ kHz\\
\br
\end{tabular}
\end{indented}
\end{table}
Typical Raman-coupled condensates therefore operate in or near $\Omega\gg\omega_x$, the locally squeezed-motional regime, rather than the $\omega\gg\Omega$ regime used to obtain a spin-only Hamiltonian.  For example, imposing $\omega_x/\Omega=100$ at $f_x=50$~Hz gives $1/\Omega\simeq0.32$~s.  The collective-mode twisting rate is $\chi=g^2/(N\omega)= (g/g_c)^2\Omega/N$, so for $N=4$ and $g/g_c=0.5$ the characteristic time $1/\chi$ is about $5.1$~s ($5.1\times10^3$~ms).  Thus the present spin-squeezing construction does not predict a realistic millisecond squeezing window for these bulk-BEC parameters.  The experiment-specific coherence time would have to exceed seconds.  By contrast, motional-quench signatures evolve on the $0.3$--$20$~ms recoil and trap scales.

\subsection{Temperature, anisotropy, and mitigation}
For the non-interacting model, a thermal product state remains separable because the evolution is local.  Thermal occupation nevertheless reduces spin contrast after tracing over motion and therefore increases $\xi_R^2$.  In an interacting gas, thermally occupied relative modes provide additional scattering and dephasing channels, and finite temperature shifts or melts the stripe phase~\cite{Ji2014FiniteTemperature}.  A quasi-one-dimensional description further requires the chemical potential, $k_BT$, recoil and Raman energies relevant to transverse scattering to remain below $\hbar\omega_\perp$.  Virtual transverse excitations renormalize the effective one-dimensional coupling, while real transverse occupation opens additional relative channels~\cite{Olshanii1998}.

Several strategies could reduce relative-mode contamination.  These include (i) spectrally resolved Raman driving detuned from relative transitions, (ii) tight and deliberately anharmonic confinement that separates collective and relative spectra, (iii) tuning scattering lengths with a Feshbach resonance to move relative resonances or reduce spin dependence, and (iv) individually trapped atoms or ions with mode-selective forces.  The last strategy is closely related to the M{\o}lmer--S{\o}rensen mechanism~\cite{PhysRevLett.82.1971}.  These approaches engineer a mode-selective model.  They are not properties of the unmodified bulk SOC Hamiltonian.

\section{Conclusions \& Outlook} \label{Sec:Conclusion}
We have identified the precise content and limitation of the cavity-QED analogy for harmonically trapped spin--orbit-coupled atoms.  One atom realizes the quantum Rabi model.  For $N$ non-interacting atoms, the center-of-mass mode produces a Dicke-type term only when considered in isolation.  The complete orthonormal set of relative modes cancels the induced spin--spin interaction exactly, and the original Hamiltonian factorizes into independent one-particle problems.  It can generate local spin--motion entanglement and motional squeezing but not interparticle entanglement or collective spin squeezing from a separable state.

This conclusion is intentionally restricted to the exact cancellation in the non-interacting harmonic model.  Contact interactions modify relative-mode denominators and matrix elements, and spin-dependent interactions can generate squeezing directly.  They may suppress relative excitations by detuning them, enhance them near resonance, or introduce additional decoherence.  They do not, however, remove the relative-mode tower.  The exact interacting SOC condensate remains multimode, and the loss of the non-interacting cancellation does not by itself produce a Dicke model.  A one-mode Dicke Hamiltonian is justified only as an effective theory after a controlled separation of all relevant relative excitations.  Likewise, single-particle interference explains how density fringes can arise, whereas the equilibrium stripe supersolid is selected and stabilized by interactions.

Realizing a clean collective simulator requires demonstrated mode selectivity rather than simply stronger spin--orbit coupling or the presence of interactions.  Spectral engineering, anharmonic confinement, interaction tuning, and individually trapped particles provide possible routes.  A quantitative theory of the interacting regime, including the competition between interaction-shifted relative modes and collective coupling, remains an important subject for future work.

\section{Acknowledgements}
The authors would like to acknowledge Salassal, Gediminas Juzeliūnas, Thomas Busch, and Ayaka Usui for fruitful discussions. This research was funded in whole or in part by the Austrian Science Fund (FWF) [10.55776/COE1].

\section*{Data availability}
All data that support the findings of this study is openly available at \cite{Hasan_zenodo:2026}.

\appendix

\section{Jacobi coordinate transformation}

We first introduce the Jacobi transformation for $N$ bosonic modes.
Consider the quadratic Hamiltonian
\begin{align}
    \hat H = \sum_{i=1}^{N} \omega \hat a_i^\dagger \hat a_i ,
\end{align}
where the operators satisfy $[\hat a_i,\hat a_j^\dagger]=\delta_{ij}$. We define one collective (center-of-mass) mode
\begin{align}
    \hat P = \frac{1}{\sqrt{N}} \sum_{i=1}^{N} \hat a_i ,
\end{align}
and $N-1$ relative (Jacobi) modes
\begin{align}
    \hat Q_k =
    \frac{1}{\sqrt{k(k+1)}}
    \left(
        \sum_{i=1}^{k} \hat a_i
        -
        k \hat a_{k+1}
    \right),
    \qquad
    k=1,\dots,N-1 .
\end{align}
The transformation is orthonormal and preserves the bosonic
commutation relations,
\begin{align}
    [\hat P,\hat P^\dagger] = 1,
    \qquad
    [\hat Q_k,\hat Q_l^\dagger] = \delta_{kl},
    \qquad
    [\hat P,\hat Q_k^\dagger] = 0 .
\end{align}

The inverse transformation reads
\begin{align}
    \hat a_j
    =
    \frac{\hat P}{\sqrt{N}}
    +
    \sum_{k=1}^{N-1}
    U_{jk}\,\hat Q_k ,
    \qquad j=1,\dots,N ,
\end{align}
with
\begin{align}
    U_{jk}
    =
    \begin{cases}
        \displaystyle \frac{1}{\sqrt{k(k+1)}} ,
        & j \le k, \\[8pt]
        \displaystyle -\frac{k}{\sqrt{k(k+1)}} ,
        & j = k+1, \\[8pt]
        0 ,
        & j > k+1 .
    \end{cases}
\end{align}
Under this transformation the quadratic Hamiltonian diagonalizes as
\begin{align}
    \sum_{i=1}^{N} \hat a_i^\dagger \hat a_i
    =
    \hat P^\dagger \hat P
    +
    \sum_{k=1}^{N-1}
    \hat Q_k^\dagger \hat Q_k .
\end{align}

\section{Four spin-orbit coupled atoms in a harmonic trap}

As a concrete illustration of the general Jacobi transformation
introduced above, we now consider $N=4$ spin-orbit coupled atoms. The SOC Hamiltonian reads
\begin{align}\label{eq:H_four_atoms}
    \hat H =
    \sum_{i=1}^{4}
    \left[
        \omega \hat a_i^\dagger \hat a_i
        +
        \frac{\Omega}{2}\hat \sigma_z^{(i)}
        +
        \frac{g}{2}\hat \sigma_x^{(i)}
        \left(\hat a_i + \hat a_i^\dagger\right)
    \right].
\end{align}
For $N=4$, the collective and relative modes defined in the previous
section reduce to
\begin{align}
    \hat P &= \frac{1}{\sqrt{4}}
    \left(
        \hat a_1+\hat a_2+\hat a_3+\hat a_4
    \right), \nonumber \\
    \hat Q &= \frac{1}{\sqrt{2}}
    \left(
        \hat a_1 - \hat a_2
    \right), \nonumber \\
    \hat R &= \frac{1}{\sqrt{6}}
    \left(
        \hat a_1+\hat a_2-2\hat a_3
    \right), \nonumber \\
    \hat S &= \frac{1}{2\sqrt{3}}
    \left(
        \hat a_1+\hat a_2+\hat a_3-3\hat a_4
    \right).
\end{align}

The inverse transformation follows directly from the general formula,
\begin{align}
    \hat a_1 &= \frac{\hat P}{2}
    + \frac{\hat Q}{\sqrt{2}}
    + \frac{\hat R}{\sqrt{6}}
    + \frac{\hat S}{2\sqrt{3}}, \nonumber \\
    \hat a_2 &= \frac{\hat P}{2}
    - \frac{\hat Q}{\sqrt{2}}
    + \frac{\hat R}{\sqrt{6}}
    + \frac{\hat S}{2\sqrt{3}}, \nonumber \\
    \hat a_3 &= \frac{\hat P}{2}
    - \frac{2\hat R}{\sqrt{6}}
    + \frac{\hat S}{2\sqrt{3}}, \nonumber \\
    \hat a_4 &= \frac{\hat P}{2}
    - \frac{3\hat S}{2\sqrt{3}}.
    \label{eq:reverse_trans_four_particles}
\end{align}
Substituting Eq.~\eqref{eq:reverse_trans_four_particles}
into Eq.~\eqref{eq:H_four_atoms}, we obtain
\begin{align}
    \hat H =\;&
    \omega
    \left(
        \hat P^\dagger \hat P
        +
        \hat Q^\dagger \hat Q
        +
        \hat R^\dagger \hat R
        +
        \hat S^\dagger \hat S
    \right)
    \nonumber \\
    &+
    \frac{\Omega}{2}
    \sum_{i=1}^{4}
    \hat \sigma_z^{(i)}
    \nonumber \\
    &+
    \frac{g}{2\sqrt{4}}
    \left(
        \hat P^\dagger + \hat P
    \right)
    \left(
        \sum_{i=1}^{4}
        \hat \sigma_x^{(i)}
    \right)
    \nonumber \\
    &+
    \frac{g}{2\sqrt{2}}
    \left(
        \hat Q^\dagger + \hat Q
    \right)
    \left(
        \hat \sigma_x^{(1)} - \hat \sigma_x^{(2)}
    \right)
    \nonumber \\
    &+
    \frac{g}{2\sqrt{6}}
    \left(
        \hat R^\dagger + \hat R
    \right)
    \left(
        \hat \sigma_x^{(1)}
        + \hat \sigma_x^{(2)}
        - 2\hat \sigma_x^{(3)}
    \right)
    \nonumber \\
    &+
    \frac{g}{4\sqrt{3}}
    \left(
        \hat S^\dagger + \hat S
    \right)
    \left(
        \hat \sigma_x^{(1)}
        + \hat \sigma_x^{(2)}
        + \hat \sigma_x^{(3)}
        - 3\hat \sigma_x^{(4)}
    \right).
    \label{eq:four_atoms_collective_relative}
\end{align}

\newpage
\section*{References}
\bibliographystyle{iopart-num}
\bibliography{ref_revised}

\end{document}